\pdfoutput=1

\documentclass[aip,jcp,amsmath,amssymb,reprint]{revtex4-1}

\usepackage{dcolumn}
\usepackage{color}

\DeclareMathOperator*{\argmin}{\arg\,\min}

\begin{document}

\newcommand*{\cion}[1]{{\text{#1}^+}}
\newcommand*{\aion}[1]{{\text{#1}^-}}
\newcommand*{\params}{{\mathcal{P}}}
\newcommand*{\interionicdistance}{{I\!D}}
\newcommand*{\latticeenergy}{{L\!E}}


\title{Crystal lattice properties fully determine short-range interaction parameters for alkali and halide ions}



\affiliation{Medical Scientist Training Program}
\affiliation{Computational and Molecular Biophysics Program, Division of Biology and Biomedical Sciences}
\affiliation{Department of Biomedical Engineering}
\affiliation{Washington University in St. Louis, One Brookings Drive, Campus Box 1097, St. Louis, MO 63130-4899}
\author{Albert H. \surname{Mao}}
\email{albert.mao@gmail.com}
\affiliation{Medical Scientist Training Program}
\affiliation{Computational and Molecular Biophysics Program, Division of Biology and Biomedical Sciences}
\affiliation{Washington University in St. Louis, One Brookings Drive, Campus Box 1097, St. Louis, MO 63130-4899}
\author{Rohit V. \surname{Pappu}}
\email{pappu@wustl.edu}
\altaffiliation[Phone: ]{(314) 935-7958}
\affiliation{Computational and Molecular Biophysics Program, Division of Biology and Biomedical Sciences}
\affiliation{Department of Biomedical Engineering}
\affiliation{Washington University in St. Louis, One Brookings Drive, Campus Box 1097, St. Louis, MO 63130-4899}

\date{\today}

\begin{abstract} 
Accurate models of alkali and halide ions in aqueous solution are necessary for computer simulations of a broad variety of systems. Previous efforts to develop ion force fields have generally focused on reproducing experimental measurements of aqueous solution properties such as hydration free energies and ion-water distribution functions. This dependency limits transferability of the resulting parameters because of the variety and known limitations of water models. We present a solvent-independent approach to calibrating ion parameters based exclusively on crystal lattice properties. Our procedure relies on minimization of lattice sums to calculate lattice energies and interionic distances instead of equilibrium ensemble simulations of dense fluids. The gain in computational efficiency enables simultaneous optimization of all parameters for Li+, Na+, K+, Rb+, Cs+, F-, Cl-, Br-, and I- subject to constraints that enforce consistency with periodic table trends. We demonstrate the method by presenting lattice-derived parameters for the primitive model and the Lennard-Jones model with Lorentz-Berthelot mixing rules. The resulting parameters successfully reproduce the lattice properties used to derive them and are free from the influence of any water model. To assess the transferability of the Lennard-Jones parameters to aqueous systems, we used them to estimate hydration free energies and found that the results were in quantitative agreement with experimentally measured values. These lattice-derived parameters are applicable in simulations where coupling of ion parameters to a particular solvent model is undesirable. The simplicity and low computational demands of the calibration procedure make it suitable for parametrization of crystallizable ions in a variety of force fields.

This article appeared in The Journal of Chemical Physics and may be accessed via its digital object identifier (DOI) name \texttt{10.1063/1.4742068}.
Copyright 2012 American Institute of Physics.
This article may be downloaded for personal use only.
Any other use requires prior permission of the author and the American Institute of Physics.
\end{abstract}




\maketitle 

\section{Introduction}

Alkali and halide ions play important roles in biological and physico-chemical systems that include protein~\cite{Roux2005Ion, Lund2008Specific, Varma2008KNa, Bostick2007Selectivity}, nucleic acid~\cite{HeilmanMiller2001Role, Chu2008Repulsive, Chen2009Molecular, Leipply2009IonRNA, Wong2010Electrostatics, Yoo2011Improved}, lipid~\cite{Cordomi2009Effect}, and carbohydrate~\cite{Eriksson2008Differential} solutions~\cite{Record1978Thermodynamic}, salt crystals~\cite{Mucha2003Salt}, molten salts~\cite{Baranyai1986Monte}, electrolytes~\cite{Fennell2009Ion}, and liquid-vapor interfaces~\cite{Netz2012Progress}.
Computer simulations are useful for developing a molecular scale description and understanding of electrolyte dependencies and ion-mediated interactions in these systems.
Most classical simulation approaches to modeling alkali and halide ions employ the Born-Oppenheimer approximation where the ions are hard spheres or van der Waals spheres with a charge of $\pm e$.
van der Waals interactions are commonly modeled using the empirical Lennard-Jones 12-6 potential.
These models for ions are used with either explicit~\cite{Barthel2002Physical} or continuum~\cite{Watanasiri1982Prediction,Vitalis2004ISIM,Lenart2007Effective,Vitalis2009Absinth} (implicit) descriptions of the surrounding solvent.
In order to achieve accuracy in simulation results, one needs reliable hard sphere or van der Waals parameters.

Experiments that measure structural and excess thermodynamic properties of electrolyte solutions can provide constraints for the calibration of these parameters.
Numerous collections of parameters have been developed for ions in aqueous solutions where the relevant constraints come from gas phase ion-water binding energies and geometries~\cite{Lee1996Molecular, Lamoureux2006Absolute, Jensen2006Halide, Joung2008Determination}, hydration free energies~\cite{Aaqvist1990IonWater, Lamoureux2006Absolute, Jensen2006Halide, Joung2008Determination, Horinek2009Rational} and entropies~\cite{Horinek2009Rational}, structural properties such as water-ion pair distribution functions~\cite{Aaqvist1990IonWater, Jensen2006Halide, Horinek2009Rational, Gee2011KirkwoodBuff}, and transport properties regarding the degree of hydration and ion association~\cite{Chen2007Quantitative}.
The resulting parameters are intricately dependent on the water model used in the parameterization procedure.
This dependence is awkward because water models themselves are more complicated and require more parameters than any one alkali or halide ion.
The need for transferable and generally applicable parameters is prominent in applications such as biomolecular simulation, where matter besides water and ions is present.
Given the sheer number, diversity, and known limitations~\cite{Vega2011Simulating} of available water models, it is difficult to be confident that ion parameters derived using a particular water model reflect the intrinsic properties of the ions that are also transferable for use in a specific simulation system.

The interionic distance and lattice energy of alkali halide salt crystals are properties that do not require any consideration of the specific model used for solvent molecules.
They constitute a set of measurable observables that constrain the length and energy scales for van der Waals interactions of non-polarizable alkali and halide ions.
In this work, we simultaneously obtain values for the sphere diameter ($\sigma$) and well depth ($\epsilon$) parameters of five alkali cations ($\cion{Li}$, $\cion{Na}$, $\cion{K}$, $\cion{Rb}$, $\cion{Cs}$) and four halide anions ($\aion{F}$, $\aion{Cl}$, $\aion{Br}$, $\aion{I}$).
These parameters, which are designed for use with a hard sphere or Lennard-Jones 12-6 potential based on Lorentz-Berthelot mixing rules, were obtained using lattice properties as the only calibration targets and are therefore independent of any specific water or solvent model.
Our calibration procedure relies on minimization of lattice sums to compute the lattice energies and interionic distances.
It requires modest computational resources compared to approaches involving explicit construction and simulation of periodic crystals or dense fluids.
We assess the transferability of the derived parameters by determining minimum energy lattice configurations and testing the accuracy of single-ion hydration free energies across three water models estimated using bicubic surfaces constructed by Joung and Cheatham~\cite{Joung2008Determination}.

\section{Methods}

The fitting is accomplished through minimization of a calibration objective function that maps any candidate parameter set to one real number quantifying deviation from experimental measurements of lattice observables.
Each evaluation of the objective function itself involves minimization of every salt's parameter-dependent potential energy to calculate its lattice energy and interionic distance.
Since this minimization does not account for thermal fluctuations, it is a ground state calculation suitable for comparison with experimental lattice energies and interionic distances measured at absolute zero.
The following sections describe these steps in detail.

\subsection{Calibration targets}

Experimental data for the twenty alkali halide salts arising from combinations of $\cion{Li}$, $\cion{Na}$, $\cion{K}$, $\cion{Rb}$, or $\cion{Cs}$ with $\aion{F}$, $\aion{Cl}$, $\aion{Br}$, or $\aion{I}$ form the basis of our calibration.
All of these salts form cubic crystals whose structures are described by one interionic distance (ID), which is defined as the distance between centers of two nearest-neighbor ions of opposite charge.
The cations and anions are arranged in interpenetrating simple cubic lattices (denoted BCC, since the unit cell is body-centered) for CsCl, CsBr, and CsI and interpenetrating face centered cubic (FCC) lattices for the other seventeen salts.
Sirdeshmukh et al. have compiled X-ray diffraction measurements of interionic distances~\cite{Sirdeshmukh2001Alkali}, while Jenkins and Roobottom have gathered lattice energy (LE) measurements~\cite{Jenkins2011Lattice} derived through the Born-Fajans-Haber thermochemical correlation~\cite{Morris1969BornFajansHaber}.
Ghate~\cite{Ghate1965ThirdOrder} extrapolated interionic distances to 0~K, a reduction of $\sim$1\% relative to room temperature values. 
The lattice energies have even lower temperature sensitivity, changing by less than 0.1\% from room temperature to 0~K~\cite{Tosi1964Cohesion}.
Therefore, we treat all parameters as temperature-independent quantities and adopt the extrapolated interionic distances and room temperature lattice energies, listed in Table~\ref{measurementtable}, as our calibration targets because they closely approximate the 0~K values that result from minimizing the potential energy of a crystal lattice.
Since there are forty independent measurements and nine ions, models with four or fewer parameters per ion are overdetermined, which is a desirable characteristic in discouraging overfitting and promoting transferability.

\begin{table}[htbp]
\caption{\label{measurementtable} Lattice energy (LE) and interionic distance (ID) measurements used as calibration targets. For each salt, the top number is the negative lattice energy and the bottom number is the interionic distance.}
\begin{ruledtabular}
\begin{tabular}{r | d d d d}
${}^{-\latticeenergy \text{ (kcal/mol)}}_{\quad \interionicdistance \text{ (\AA{})}}$ & \multicolumn{1}{c}{F} & \multicolumn{1}{c}{Cl} & \multicolumn{1}{c}{Br} & \multicolumn{1}{c}{I} \\
\hline
Li & 250.7 & 206.5 & 196.0 & 182.6 \\
 & 1.996 & 2.539 & 2.713 & 2.951 \\
Na & 222.3 & 188.8 & 180.2 & 168.5 \\
 & 2.295 & 2.789 & 2.954 & 3.194 \\
K & 198.1 & 172.1 & 165.2 & 155.4 \\
 & 2.648 & 3.116 & 3.262 & 3.489 \\
Rb & 190.0 & 166.1 & 159.7 & 151.1 \\
 & 2.789 & 3.259 & 3.410 & 3.628 \\
Cs & 181.4 & 160.1 & 154.6 & 146.5 \\
 & 2.982 & 3.523 & 3.668 & 3.898
\end{tabular}
\end{ruledtabular}
\end{table}

\subsection{Ion interaction models}

We focus on two common models for ions in molecular simulations.
Both models are pairwise additive potentials where a short-range interaction is superimposed upon the Coulomb electrostatic interaction.
For a pair of particles denoted as $i$ and $j$,
\begin{equation} \label{eq:coulomb}
U^{\text{elec}}_{ij}(r_{ij}) \equiv \frac{k z_i z_j}{r_{ij}}
\end{equation}
where $k \equiv e^2 / 4 \pi \epsilon_0 \approx 332.06\text{ kcal/mol}$, the valences $z$ are 1 for alkali cations and $-1$ for halide anions, and $r_{ij}$ is the distance between their centers.
The dielectric constant is uniformly 1 because the salt crystals are modeled in the absence of other matter.

In the primitive model (PM), ions are hard spheres which cannot overlap.
Each ion has one parameter, its diameter $\sigma$:
\begin{equation} \label{eq:primitivemodel}
U^{\text{PM}}_{ij}(r_{ij}) \equiv U^{\text{elec}}_{ij}(r_{ij}) + 
\begin{cases}
\infty &\text{if } r_{ij} < \frac{1}{2}(\sigma_i + \sigma_j) \\
0 &\text{otherwise}
\end{cases}
\end{equation}

In the Lennard-Jones (LJ) model, ions exhibit short-range attractive van der Waals interactions that compete against a repulsive barrier.
Each ion has two parameters, $\sigma$ and $\epsilon$, which respectively describe the length and energy scales of its interactions.
We adopt Lorentz-Berthelot mixing rules, which specify arithmetic means for $\sigma$ and geometric means for $\epsilon$, to combine the parameters of two ions into one pairwise interaction.

\begin{equation} \label{eq:mixingrules}
\sigma_{ij} \equiv \frac{1}{2} ( \sigma_{i} + \sigma_{j} ) \qquad \epsilon_{ij} \equiv \sqrt{\epsilon_{i} \epsilon_{j}}
\end{equation}
\begin{equation} \label{eq:lennardjonesmodel}
U^{\text{LJ}}_{ij}(r_{ij}) \equiv U^{\text{elec}}_{ij}(r_{ij}) + 4 \epsilon_{ij} \left[ \left( \frac{\sigma_{ij}}{r_{ij}} \right)^{12} - \left( \frac{\sigma_{ij}}{r_{ij}} \right) ^6 \right]
\end{equation}

\subsection{Lattice energy and interionic distance calculations}

Due to the symmetry of a periodic alkali halide crystal lattice, the contribution from one ion to the total potential energy is the same for every cation and for every anion.
This contribution is equal to the potential energy of one ion in the field generated by all other ions in the lattice, divided by two to correct for double counting.
The sum of the contributions from one cation and one anion, respectively denoted as $c$ and $a$, gives the potential energy per salt pair as a function of the tentative interionic distance $d$:
\begin{equation} \label{eq:intensivepotentialenergy}
U_{ca}(d) \equiv \frac{1}{2} \sum_i^{\{c,a\}} \sum_{j \ne i}^{\{\text{lattice}\}} U_{ij}(r_{ij}(d))
\end{equation}

Proceeding in a manner analogous to the derivation of the classic Born-Land\'{e} equation, we compute the calibration observables by minimizing this intensive potential energy with respect to $d$.
The minimum energy is the lattice energy ($\latticeenergy$) and optimal value of $d$ is the interionic distance ($\interionicdistance$).
Since the shape of $U_{ca}(d)$ depends on the parameters $\params$, the calibration observables are functions of $\params$.
In Equations~\ref{eq:calibrationobservables}, \ref{eq:pmobservables}, and \ref{eq:ljobservables}, we make this functional relation explicit.
$\params_{ca}$ denotes parameters pertaining to the cation $c$ and anion $a$ and consists of the pair $(\sigma_c, \sigma_a)$ for the primitive model and the quadruple $(\sigma_c, \epsilon_c, \sigma_a, \epsilon_a)$ for the Lennard-Jones model.
\begin{align} \label{eq:calibrationobservables}
\interionicdistance_{ca}(\params_{ca}) &\equiv \argmin_{d > 0} U_{ca}(d) \nonumber \\
\latticeenergy_{ca}(\params_{ca}) &\equiv \min_{d > 0} U_{ca}(d)
\end{align}

For the primitive model, since the Coulomb interaction draws the lattice together as tightly as possible, the optimal value of $d$ is the one at which lattice ions come into contact with each other:
\begin{align} \label{eq:pmobservables}
&U^\text{PM}_{ca}(d) = 
\begin{cases}
\infty &\text{if } d < \max \left\{ s \sigma_c, s \sigma_a, \frac{1}{2}(\sigma_c + \sigma_a) \right\} \\
-\frac{b_1 k}{d} &\text{otherwise}
\end{cases} \nonumber \\
&\interionicdistance^\text{PM}_{ca}(\sigma_c, \sigma_a) = \max \left\{ s \sigma_c, s \sigma_a, \frac{1}{2}(\sigma_c + \sigma_a) \right\} \nonumber \\
&\latticeenergy^\text{PM}_{ca}(\sigma_c, \sigma_a) = -\frac{b_1 k}{\interionicdistance^\text{PM}_{ca}(\sigma_c, \sigma_a)}
\end{align}
where $s$ is $1/\sqrt{2}$ for FCC lattices and $\sqrt{3}/2$ for BCC lattices and $b_1$, the Madelung constant, is approximately 1.7476 for FCC lattices and 1.7627 for BCC lattices.

For the Lennard-Jones model, the intensive potential energy can be written as a rational function of $d$ because the lattice sums $b_{12}$ and $b_6$ of the Lennard-Jones potential are numerical constants and can be precomputed:
\begin{align} \label{eq:ljenergy}
U^\text{LJ}_{ca}(d) \equiv &- \frac{b_1 k}{d} \nonumber \\
&+ \sum_{i,j}^{\{c,a\}}  2 \epsilon_{ij} \left[ b_{12,ij} \left( \frac{\sigma_{ij}}{d} \right)^{12} - b_{6,ij} \left( \frac{\sigma_{ij}}{d} \right) ^6 \right]
\end{align}
Unlike the classic Madelung constant, these lattice sums are absolutely convergent~\cite{Borwein1985Convergence}.
They can be approximated by summing over a finite cube centered at and omitting the origin, but care must be taken to prevent inexact floating point arithmetic from making the summation converge to an inaccurate result~\cite{Higham1993Accuracy}.
We use exact rational arithmetic for accumulating these sums and convert the final totals to IEEE 754 double-precision floating point numbers.
Double-precision floating point arithmetic is used for all other numerical operations in this study.
We sum over a cube with $601^3$ ions, keeping separate totals for ``even'' and ``odd'' lattice sites, to obtain the numerical values for FCC lattices in Equation~\ref{eq:fccmadelung}:
\begin{align} \label{eq:fccmadelung}
b_{6,cc} = b_{6,aa} \approx 1.8067 \quad b_{6,ca} = b_{6,ac} \approx 6.5952 \nonumber \\
b_{12,cc} = b_{12,aa} \approx 0.1896 \quad b_{12,ca} = b_{12,ac} \approx 6.0126
\end{align}
Likewise, we perform separate sums over cubes with $601^3$ and $600^3$ ions for ``like'' and ``unlike'' lattice sites, respectively, to obtain the corresponding BCC lattice sums.
Equation~\ref{eq:bccmadelung} presents these sums multiplied by $\sqrt{3}/2$, the ratio of interionic distance to lattice constant, raised to the sixth power for $b_6$ or twelfth power for $b_{12}$:
\begin{align} \label{eq:bccmadelung}
b_{6,cc} = b_{6,aa} \approx 3.5446 \quad b_{6,ca} = b_{6,ac} \approx 8.7091 \nonumber \\
b_{12,cc} = b_{12,aa} \approx 1.1038 \quad b_{12,ca} = b_{12,ac} \approx 8.0103
\end{align}
Decreasing cube side lengths by a factor of three changes the resulting sums by an amount less than $1.2 \times 10^{-6}$.

Minimizing $U^{\text{LJ}}_{ca}(d)$ requires the real positive roots of the polynomial $Q$ in Equation~\ref{eq:ljpolynomial}, which was obtained by implementing the mixing rules in Equation~\ref{eq:mixingrules} and factoring the derivative $\mathrm{d} U^{\text{LJ}}_{ca} / \mathrm{d}d$:
\begin{align} \label{eq:ljpolynomial}
&Q_{ca}(d) = -256 b_1 k d^{11} \nonumber \\
&\quad - {} 96 \left[ b_{6,ca} \sqrt{\epsilon_c \epsilon_a} \left( \sigma_c + \sigma_a \right)^6 \right. \nonumber \\
&\qquad + {} 32 \left. \left( b_{6,cc} \epsilon_c \sigma_c^6 + b_{6,aa} \epsilon_a \sigma_a^6 \right) \right] d^6 \nonumber \\
&\quad + {} 3 \left[ b_{12,ca} \sqrt{\epsilon_c \epsilon_a} \left( \sigma_c + \sigma_a \right)^{12} \right. \nonumber \\
&\qquad + {} 2048 \left. \left( b_{12,cc} \epsilon_c \sigma_c^{12} + b_{12,aa} \epsilon_a \sigma_a^{12} \right) \right]
\end{align}
Given particular numerical values for the parameters $\sigma_c$, $\epsilon_c$, $\sigma_a$, and $\epsilon_a$, lattice sums $b$, and electrostatic constant $k$, $Q$ becomes an eleventh-degree polynomial in $d$ whose roots can be obtained via standard methods such as the Jenkins-Traub algorithm~\cite{Jenkins1970ThreeStage}.
In cases where the derivative has multiple positive roots, the smallest one is taken to define the interionic distance:
\begin{align} \label{eq:ljobservables}
\interionicdistance^\text{LJ}_{ca}(\sigma_c, \epsilon_c, \sigma_a, \epsilon_a) &= \min \left\{ d \in \mathbb{R}_{>0} : Q_{ca}(d)=0 \right\} \nonumber \\
\latticeenergy^\text{LJ}_{ca}(\sigma_c, \epsilon_c, \sigma_a, \epsilon_a) &= U^\text{LJ}_{ca} \left( \interionicdistance^\text{LJ}_{ca}({\sigma_c, \epsilon_c, \sigma_a, \epsilon_a}) \right)
\end{align}
If no positive roots exist, the situation corresponds to an unstable crystal with undefined calibration observables.

\subsection{Calibration objective function and parameter constraints}

The root mean square relative deviation from $O^\text{measured}$, the measured values of calibration targets given in Table~\ref{measurementtable}, is used as the objective function whose minimization yields the optimized parameters.
\begin{equation} \label{eq:objectivefunction}
F(\params) \equiv \sqrt{ \frac{1}{40} \sum_{ca}^{\{\text{salts}\}} \sum_O^{\{\interionicdistance,\latticeenergy\}} \left( \frac{O_{ca}(\params_{ca})}{O_{ca}^\text{measured}} - 1 \right)^2 }
\end{equation}
In Equation~\ref{eq:objectivefunction}, the outer sum is over all twenty cation-anion salt pairs, and the inner sum is over the two observables $\interionicdistance$ and $\latticeenergy$.
We used relative instead of absolute deviations because they put all the observables on an equal footing; biases due to differing magnitudes and units are naturally eliminated.
As a special case, parameters that cause any salt crystal to be unstable are defined to have an infinite calibration objective function value.
The domain $\{\params\}$ of the function consists of the nine-dimensional space $\{\sigma\}^9$ for the primitive model and the eighteen-dimensional space $\{\sigma, \epsilon\}^9$ for the Lennard-Jones model.

Following the reasoning of Peng et al.~\cite{Peng1997Derivation}, we constrain the domain to maintain consistency with periodic table trends.
Ions increase in size going down their respective groups of the periodic table, and cations are smaller than their isoelectronic anions.
For both the primitive and Lennard-Jones models, Equations~\ref{eq:sigmaconstraints1}~and~\ref{eq:sigmaconstraints2} show the constraints applied to $\sigma$, which correspond to the ion diameters:
\begin{align} \label{eq:sigmaconstraints1}
0 &< \sigma_\cion{Li} < \sigma_\cion{Na} < \sigma_\cion{K} < \sigma_\cion{Rb} < \sigma_\cion{Cs} \nonumber \\
0 &< \sigma_\aion{F} < \sigma_\aion{Cl} < \sigma_\aion{Br} < \sigma_\aion{I}
\end{align}
\begin{align} \label{eq:sigmaconstraints2}
\sigma_\cion{Na} &< \sigma_\aion{F} \nonumber \\
\sigma_\cion{K} &< \sigma_\aion{Cl} \nonumber \\
\sigma_\cion{Rb} &< \sigma_\aion{Br} \nonumber \\
\sigma_\cion{Cs} &< \sigma_\aion{I}
\end{align}
In accord with the ions' isoelectronic noble gases, the Lennard-Jones well depths $\epsilon$ also increase going down each group (Equation~\ref{eq:epsilonconstraints}):
\begin{align} \label{eq:epsilonconstraints}
0 &< \epsilon_\cion{Li} < \epsilon_\cion{Na} < \epsilon_\cion{K} < \epsilon_\cion{Rb} < \epsilon_\cion{Cs} \nonumber \\
0 &< \epsilon_\aion{F} < \epsilon_\aion{Cl} < \epsilon_\aion{Br} < \epsilon_\aion{I}
\end{align}
The scale of the London dispersion interaction, which corresponds to the coefficient of the $r^{-6}$ term in the overall potential, is smaller for cations compared to isoelectronic anions because of their lower polarizabilities.
It follows that this coefficient, which is equal to $4 \epsilon \sigma^6$ in the Lennard-Jones model, must obey the inequalities in Equation~\ref{eq:csixconstraints}:
\begin{align} \label{eq:csixconstraints}
\epsilon_\cion{Na}(\sigma_\cion{Na})^6  &< \epsilon_\aion{F} (\sigma_\aion{F})^6 \nonumber \\
\epsilon_\cion{K} (\sigma_\cion{K})^6 &< \epsilon_\aion{Cl} (\sigma_\aion{Cl})^6 \nonumber \\
\epsilon_\cion{Rb} (\sigma_\cion{Rb})^6 &< \epsilon_\aion{Br} (\sigma_\aion{Br})^6 \nonumber \\
\epsilon_\cion{Cs} (\sigma_\cion{Cs})^6 &< \epsilon_\aion{I} (\sigma_\aion{I})^6
\end{align}
Imposition of these constraints focuses the search on the subset of parameter space that is physically reasonable and promotes transferability of the resulting parameters.

\subsection{Constrained nonlinear numerical optimization}

We implemented the interaction models and calibration objective function with Mathematica~7 (Wolfram Research) and used its constrained nonlinear numerical optimization routines to simultaneously determine all parameters through minimization of the objective function.
Initial trials showed that for this global optimization problem, the differential evolution method~\cite{Storn1997Differential} achieved better performance than simulated annealing~\cite{Kirkpatrick1983Optimization}, the Nelder-Mead simplex method~\cite{Nelder1965Simplex}, and local minimization from random initial points.
Differential evolution is an iterative general-purpose function minimizer that evolves a population of solutions to search for the global minimum.
In our application, each solution is a set of parameters for all nine ions.
During a single iteration, every member of the population competes against a perturbed version of itself for survival.
Perturbations consist of crossing a mutant parameter set with the original such that each parameter randomly inherits its value from either the mutant or the original.
Mutant sets are generated by randomly selecting three distinct members of the population and vectorially adding the first to a scaled difference of the other two.
If the perturbed solution improves the objective function score, it replaces the original in the population for the next iteration.
The crossover and mutation processes make differential evolution robust in the presence of many local minima and do not require evaluation of objective function gradients.

In employing differential evolution, we used a population size of 100, preserved the default options of 0.5 for the cross probability and 0.6 for the scaling factor, and enabled solution post-processing by the interior point local minimization algorithm~\cite{Wright2004Interiorpoint}.
In both differential evolution and local minimization, the convergence criterion was an absolute or proportional change of less than $10^{-8}$ in the appropriate units for all parameters and the objective function value.
To facilitate convergence of the search procedure, we supplied randomly generated initial guesses that satisfied most or all of the constraints given in Equations~\ref{eq:sigmaconstraints1}--\ref{eq:csixconstraints}.
For each set of initial $\sigma$ guesses, nine uniformly and independently distributed random diameters between 1~\AA{} and 5~\AA{} were generated, sorted in ascending order, and assigned to the nine ions according to a random permutation uniformly selected from the fourteen that are consistent with Equations~\ref{eq:sigmaconstraints1} and \ref{eq:sigmaconstraints2}.
For each set of initial Lennard-Jones $\epsilon$ guesses, five log-uniformly and independently distributed random interaction energies between $0.001$ and $0.75$~kcal/mol were generated, sorted, and assigned to the five cations, with a separate four to the anions, such that the constraints in Equation~\ref{eq:epsilonconstraints} were satisfied but those from Equation~\ref{eq:csixconstraints} were sometimes violated.
These bounds on the initial guesses did not limit the evolution of the parameters during the minimization process.
Initial guesses with constraint violations evolved towards compliance and still contributed during the early iterations via mutant generation.

The entire differential evolution process was executed 100 times with distinct random seeds.
Therefore, $10^4$ distinct initial guesses were evolved in independent groups of 100 to produce 100 population champions that we compared to select the optimal parameter set.
In total, each model's objective function was evaluated on the order of $5 \times 10^6$ times.
Despite the robustness of differential evolution and the large number of parameter sets evaluated, the parameter space is high dimensional and global optimality cannot be guaranteed.
For the primitive model, all 100 populations converged toward one of two solutions.
We selected the one with the better objective function score as our final recommended primitive model parameter set.
In contrast, for the Lennard-Jones model, the 100 populations each produced a distinct champion parameter set and objective function score, with many pushing tightly against the constraints.
This suggests that the function landscape is rugged with many local minima, and that deeper minima corresponding to unphysical parameters exist outside the region allowed by the constraints.
We selected the champion with the best objective function score that satisfied all constraints to a tolerance of at least $10^{-5}$ in their respective units as our final recommended Lennard-Jones parameter set.

\section{Results}

\subsection{Calibrated ion parameters}

Final optimized parameters for both the primitive model and Lennard-Jones model are presented in Table~\ref{parametertable}.
As expected, all parameters satisfy the periodic table trends expressed in Equations~\ref{eq:sigmaconstraints1}--\ref{eq:csixconstraints}.
If satisfaction of periodic table trends is desired, it is necessary to impose the constraints while simultaneously optimizing all parameters; calibration protocols that do not include these constraints are likely to produce parameters that violate them.
In some cases~\cite{Aaqvist1990IonWater, Gee2011KirkwoodBuff}, the attained Lennard-Jones $\epsilon$ actually reverse the expected trend by decreasing down each ion group.
In addition to having correct relative magnitudes, the ranges of 1.7 to 5.2~\AA{} for $\sigma$ and 0.006 to 0.5~kcal/mol for $\epsilon$ are comparable to those of Lennard-Jones parameters for noble gases~\cite{Mourits1977Critical}, even though no absolute bounds other than positivity were enforced during their optimization.

\begin{table}[htbp]
\caption{\label{parametertable} Short-range interaction parameters derived from crystal lattice properties.}
\begin{ruledtabular}
\begin{tabular}{l l l l}
Ion & Primitive model & \multicolumn{2}{l}{Lennard-Jones model} \\
 & $\sigma$ (\AA{}) & $\sigma$ (\AA{}) & $\epsilon$ (kcal/mol) \\
\hline
$\cion{Li}$ & 1.716 & 1.715 & 0.05766 \\
$\cion{Na}$ & 2.271 & 2.497 & 0.07826 \\
$\cion{K}$ & 2.902 & 3.184 & 0.1183 \\
$\cion{Rb}$ & 3.165 & 3.302 & 0.2405 \\
$\cion{Cs}$ & 3.559 & 3.440 & 0.5013 \\
\hline
$\aion{F}$ & 2.626 & 3.954 & 0.006465 \\
$\aion{Cl}$ & 3.600 & 4.612 & 0.02502 \\
$\aion{Br}$ & 3.903 & 4.812 & 0.03596 \\
$\aion{I}$ & 4.331 & 5.197 & 0.04220  
\end{tabular}
\end{ruledtabular}
\end{table}

\subsection{Attained values of calibration observables}

The optimized hard sphere diameters for the primitive model attain a relative root mean square deviation (rRMSD) from calibration targets of 4.3\%, with RMSDs for interionic distances and lattice energies being 0.12~\AA{} and 8.5~kcal/mol, respectively.
While the anions are diminished relative to their crystallographically derived ionic diameters~\cite{Shannon1976Revised}, the cations are swollen by a greater amount.
As shown in Table~\ref{pmattained}, the attained magnitudes for all lattice energies and interionic distances are greater than their measured values.
This indicates that lattice energies naively calculated using ionic radii as the hard sphere radii would be too negative, necessitating an overall swelling that trades off accuracy in interionic distances in exchange for better accuracy in lattice energies.
The resulting compromise can be considered a best fit of the primitive model to lattice data, and highlights the deficiency of hard sphere exclusion as a model for short-range repulsion between ions.
The need to introduce an overall swelling of ion diameters to match experimental measurements has also been encountered in studies of primitive model activity coefficients in the mean spherical approximation~\cite{Watanasiri1982Prediction}.

\begin{table}[htbp]
\caption{\label{pmattained} Negative lattice energies and interionic distances attained using lattice-derived primitive model parameters. Digits colored blue indicate positive deviations from calibration targets. In each number, the most significant colored digit is shaded to indicate the magnitude of the deviation, with darker shades indicating smaller deviations.}
\begin{ruledtabular}
\begin{tabular}{r | d d d d}
${}^{-\latticeenergy \text{ (kcal/mol)}}_{\quad \interionicdistance \text{ (\AA{})}}$ & \multicolumn{1}{c}{F} & \multicolumn{1}{c}{Cl} & \multicolumn{1}{c}{Br} & \multicolumn{1}{c}{I} \\
\hline
Li & 2\textcolor[rgb]{0.00, 0.00, 0.733}{6}\textcolor[rgb]{0.00, 0.00, 1.00}{7}.\textcolor[rgb]{0.00, 0.00, 1.00}{3} & 2\textcolor[rgb]{0.00, 0.00, 0.700}{1}\textcolor[rgb]{0.00, 0.00, 1.00}{8}.\textcolor[rgb]{0.00, 0.00, 1.00}{3} & 2\textcolor[rgb]{0.00, 0.00, 0.700}{0}\textcolor[rgb]{0.00, 0.00, 1.00}{6}.\textcolor[rgb]{0.00, 0.00, 1.00}{5} & 18\textcolor[rgb]{0.00, 0.00, 0.900}{9}.\textcolor[rgb]{0.00, 0.00, 1.00}{5} \\
 & 2.\textcolor[rgb]{0.00, 0.00, 0.733}{1}\textcolor[rgb]{0.00, 0.00, 1.00}{7}\textcolor[rgb]{0.00, 0.00, 1.00}{1} & 2.\textcolor[rgb]{0.00, 0.00, 0.700}{6}\textcolor[rgb]{0.00, 0.00, 1.00}{5}\textcolor[rgb]{0.00, 0.00, 1.00}{8} & 2.\textcolor[rgb]{0.00, 0.00, 0.700}{8}\textcolor[rgb]{0.00, 0.00, 1.00}{1}\textcolor[rgb]{0.00, 0.00, 1.00}{0} & 3.\textcolor[rgb]{0.00, 0.00, 0.700}{0}\textcolor[rgb]{0.00, 0.00, 1.00}{6}\textcolor[rgb]{0.00, 0.00, 1.00}{2} \\
Na & 2\textcolor[rgb]{0.00, 0.00, 0.700}{3}\textcolor[rgb]{0.00, 0.00, 1.00}{7}.\textcolor[rgb]{0.00, 0.00, 1.00}{0} & 19\textcolor[rgb]{0.00, 0.00, 0.967}{7}.\textcolor[rgb]{0.00, 0.00, 1.00}{7} & 18\textcolor[rgb]{0.00, 0.00, 0.933}{8}.\textcolor[rgb]{0.00, 0.00, 1.00}{0} & 17\textcolor[rgb]{0.00, 0.00, 0.900}{5}.\textcolor[rgb]{0.00, 0.00, 1.00}{8} \\
 & 2.\textcolor[rgb]{0.00, 0.00, 0.733}{4}\textcolor[rgb]{0.00, 0.00, 1.00}{4}\textcolor[rgb]{0.00, 0.00, 1.00}{9} & 2.\textcolor[rgb]{0.00, 0.00, 0.700}{9}\textcolor[rgb]{0.00, 0.00, 1.00}{3}\textcolor[rgb]{0.00, 0.00, 1.00}{6} & 3.\textcolor[rgb]{0.00, 0.00, 0.700}{0}\textcolor[rgb]{0.00, 0.00, 1.00}{8}\textcolor[rgb]{0.00, 0.00, 1.00}{7} & 3.\textcolor[rgb]{0.00, 0.00, 0.700}{3}\textcolor[rgb]{0.00, 0.00, 1.00}{0}\textcolor[rgb]{0.00, 0.00, 1.00}{1} \\
K & 2\textcolor[rgb]{0.00, 0.00, 0.700}{1}\textcolor[rgb]{0.00, 0.00, 1.00}{0}.\textcolor[rgb]{0.00, 0.00, 1.00}{0} & 17\textcolor[rgb]{0.00, 0.00, 0.867}{8}.\textcolor[rgb]{0.00, 0.00, 1.00}{5} & 17\textcolor[rgb]{0.00, 0.00, 0.833}{0}.\textcolor[rgb]{0.00, 0.00, 1.00}{6} & 16\textcolor[rgb]{0.00, 0.00, 0.833}{0}.\textcolor[rgb]{0.00, 0.00, 1.00}{5} \\
 & 2.\textcolor[rgb]{0.00, 0.00, 0.700}{7}\textcolor[rgb]{0.00, 0.00, 1.00}{6}\textcolor[rgb]{0.00, 0.00, 1.00}{4} & 3.\textcolor[rgb]{0.00, 0.00, 0.700}{2}\textcolor[rgb]{0.00, 0.00, 1.00}{5}\textcolor[rgb]{0.00, 0.00, 1.00}{1} & 3.\textcolor[rgb]{0.00, 0.00, 0.700}{4}\textcolor[rgb]{0.00, 0.00, 1.00}{0}\textcolor[rgb]{0.00, 0.00, 1.00}{2} & 3.\textcolor[rgb]{0.00, 0.00, 0.700}{6}\textcolor[rgb]{0.00, 0.00, 1.00}{1}\textcolor[rgb]{0.00, 0.00, 1.00}{6} \\
Rb & 2\textcolor[rgb]{0.00, 0.00, 0.700}{0}\textcolor[rgb]{0.00, 0.00, 1.00}{0}.\textcolor[rgb]{0.00, 0.00, 1.00}{4} & 17\textcolor[rgb]{0.00, 0.00, 0.833}{1}.\textcolor[rgb]{0.00, 0.00, 1.00}{6} & 16\textcolor[rgb]{0.00, 0.00, 0.833}{4}.\textcolor[rgb]{0.00, 0.00, 1.00}{2} & 15\textcolor[rgb]{0.00, 0.00, 0.800}{4}.\textcolor[rgb]{0.00, 0.00, 1.00}{8} \\
 & 2.\textcolor[rgb]{0.00, 0.00, 0.700}{8}\textcolor[rgb]{0.00, 0.00, 1.00}{9}\textcolor[rgb]{0.00, 0.00, 1.00}{5} & 3.\textcolor[rgb]{0.00, 0.00, 0.700}{3}\textcolor[rgb]{0.00, 0.00, 1.00}{8}\textcolor[rgb]{0.00, 0.00, 1.00}{3} & 3.\textcolor[rgb]{0.00, 0.00, 0.700}{5}\textcolor[rgb]{0.00, 0.00, 1.00}{3}\textcolor[rgb]{0.00, 0.00, 1.00}{4} & 3.\textcolor[rgb]{0.00, 0.00, 0.700}{7}\textcolor[rgb]{0.00, 0.00, 1.00}{4}\textcolor[rgb]{0.00, 0.00, 1.00}{8} \\
Cs & 18\textcolor[rgb]{0.00, 0.00, 0.867}{7}.\textcolor[rgb]{0.00, 0.00, 1.00}{7} & 16\textcolor[rgb]{0.00, 0.00, 0.767}{3}.\textcolor[rgb]{0.00, 0.00, 1.00}{5} & 15\textcolor[rgb]{0.00, 0.00, 0.733}{6}.\textcolor[rgb]{0.00, 0.00, 1.00}{9} & 14\textcolor[rgb]{0.00, 0.00, 0.733}{8}.\textcolor[rgb]{0.00, 0.00, 1.00}{4} \\
 & 3.\textcolor[rgb]{0.00, 0.00, 0.700}{0}\textcolor[rgb]{0.00, 0.00, 1.00}{9}\textcolor[rgb]{0.00, 0.00, 1.00}{2} & 3.5\textcolor[rgb]{0.00, 0.00, 0.867}{8}\textcolor[rgb]{0.00, 0.00, 1.00}{0} & 3.7\textcolor[rgb]{0.00, 0.00, 0.867}{3}\textcolor[rgb]{0.00, 0.00, 1.00}{1} & 3.9\textcolor[rgb]{0.00, 0.00, 0.833}{4}\textcolor[rgb]{0.00, 0.00, 1.00}{5}
\end{tabular}
\end{ruledtabular}
\end{table}

The optimized Lennard-Jones parameters attain a rRMSD from calibration targets of 1.4\%, with RMSDs for interionic distances and lattice energies being 0.031~\AA{} and 3.0~kcal/mol, respectively.
Table~\ref{ljattained} shows that these deviations are smaller and more balanced than those of the primitive model, reflecting an improved ability of the Lennard-Jones model to capture the essential physics of van der Waals interactions.
These deviations are also smaller than those obtained by Peng et al.~\cite{Peng1997Derivation}, which is expected since they did not optimize all parameters simultaneously.
Compared to Joung and Cheatham's~\cite{Joung2008Determination} recommended parameters for use with the TIP3P water model, our deviations are lower for lattice energies and higher for interionic distances.
However, the significance of these comparisons is limited due to the distinct calibration targets used in each study and the violation of periodic table trend constraints by the Joung and Cheatham parameters.
We rejected several candidate parameter sets with better objective function scores because they saturated one or more constraints by having multiple ions with nearly identical $\sigma$, $\epsilon$, and/or $r^{-6}$ coefficients.
This suggests that unconstrained minimization of the objective function would lead to even lower objective function scores, underscoring the importance of the constraints in maintaining physical realism.

\begin{table}[hbtp]
\caption{\label{ljattained} Negative lattice energies and interionic distances attained using lattice-derived Lennard-Jones parameters with Lorentz-Berthelot mixing rules. Digits colored blue/red indicate where attained values exceed/fall short of calibration targets. In each number, the most significant colored digit is shaded to indicate the magnitude of the deviation, with darker shades indicating smaller deviations.}
\begin{ruledtabular}
\begin{tabular}{r | d d d d}
${}^{-\latticeenergy \text{ (kcal/mol)}}_{\quad \interionicdistance \text{ (\AA{})}}$ & \multicolumn{1}{c}{F} & \multicolumn{1}{c}{Cl} & \multicolumn{1}{c}{Br} & \multicolumn{1}{c}{I} \\
\hline
Li & 25\textcolor[rgb]{0.00, 0.00, 0.933}{8}.\textcolor[rgb]{0.00, 0.00, 1.00}{9} & 21\textcolor[rgb]{0.00, 0.00, 0.800}{0}.\textcolor[rgb]{0.00, 0.00, 1.00}{7} & 19\textcolor[rgb]{0.00, 0.00, 0.733}{8}.\textcolor[rgb]{0.00, 0.00, 1.00}{4} & 182.5 \\
 & 2.0\textcolor[rgb]{0.00, 0.00, 0.933}{7}\textcolor[rgb]{0.00, 0.00, 1.00}{3} & 2.5\textcolor[rgb]{0.00, 0.00, 0.767}{6}\textcolor[rgb]{0.00, 0.00, 1.00}{5} & 2.7\textcolor[rgb]{0.00, 0.00, 0.733}{3}\textcolor[rgb]{0.00, 0.00, 1.00}{2} & 2.9\textcolor[rgb]{0.00, 0.00, 0.733}{7}\textcolor[rgb]{0.00, 0.00, 1.00}{6} \\
Na & 22\textcolor[rgb]{0.00, 0.00, 0.800}{6}.\textcolor[rgb]{0.00, 0.00, 1.00}{5} & 19\textcolor[rgb]{0.00, 0.00, 0.767}{1}.\textcolor[rgb]{0.00, 0.00, 1.00}{7} & 18\textcolor[rgb]{0.00, 0.00, 0.767}{2}.\textcolor[rgb]{0.00, 0.00, 1.00}{9} & 17\textcolor[rgb]{0.00, 0.00, 0.733}{0}.\textcolor[rgb]{0.00, 0.00, 1.00}{8} \\
 & 2.3\textcolor[rgb]{0.00, 0.00, 0.933}{7}\textcolor[rgb]{0.00, 0.00, 1.00}{3} & 2.8\textcolor[rgb]{0.00, 0.00, 0.767}{2}\textcolor[rgb]{0.00, 0.00, 1.00}{2} & 2.9\textcolor[rgb]{0.00, 0.00, 0.700}{6}\textcolor[rgb]{0.00, 0.00, 1.00}{8} & 3.18\textcolor[rgb]{0.967, 0.00, 0.00}{5} \\
K & 20\textcolor[rgb]{0.00, 0.00, 0.767}{0}.\textcolor[rgb]{0.00, 0.00, 1.00}{7} & 172.\textcolor[rgb]{0.00, 0.00, 0.933}{9} & 166.\textcolor[rgb]{0.00, 0.00, 0.933}{0} & 15\textcolor[rgb]{0.00, 0.00, 0.700}{6}.\textcolor[rgb]{0.00, 0.00, 1.00}{6} \\
 & 2.6\textcolor[rgb]{0.00, 0.00, 0.800}{8}\textcolor[rgb]{0.00, 0.00, 1.00}{6} & 3.1\textcolor[rgb]{0.00, 0.00, 0.733}{3}\textcolor[rgb]{0.00, 0.00, 1.00}{7} & 3.2\textcolor[rgb]{0.00, 0.00, 0.733}{7}\textcolor[rgb]{0.00, 0.00, 1.00}{8} & 3.48\textcolor[rgb]{0.933, 0.00, 0.00}{1} \\
Rb & 19\textcolor[rgb]{0.00, 0.00, 0.767}{2}.\textcolor[rgb]{0.00, 0.00, 1.00}{7} & 166.\textcolor[rgb]{0.00, 0.00, 0.767}{4} & 160.\textcolor[rgb]{0.00, 0.00, 0.767}{0} & 151.\textcolor[rgb]{0.00, 0.00, 0.733}{2} \\
 & 2.8\textcolor[rgb]{0.00, 0.00, 0.733}{1}\textcolor[rgb]{0.00, 0.00, 1.00}{4} & 3.2\textcolor[rgb]{0.00, 0.00, 0.733}{7}\textcolor[rgb]{0.00, 0.00, 1.00}{5} & 3.41\textcolor[rgb]{0.00, 0.00, 0.900}{7} & 3.62\textcolor[rgb]{0.867, 0.00, 0.00}{2} \\
Cs & 18\textcolor[rgb]{0.00, 0.00, 0.800}{5}.\textcolor[rgb]{0.00, 0.00, 1.00}{4} & 15\textcolor[rgb]{0.700, 0.00, 0.00}{8}.\textcolor[rgb]{1.00, 0.00, 0.00}{8} & 15\textcolor[rgb]{0.733, 0.00, 0.00}{2}.\textcolor[rgb]{1.00, 0.00, 0.00}{5} & 14\textcolor[rgb]{0.767, 0.00, 0.00}{3}.\textcolor[rgb]{1.00, 0.00, 0.00}{8} \\
 & 2.9\textcolor[rgb]{0.767, 0.00, 0.00}{5}\textcolor[rgb]{1.00, 0.00, 0.00}{3} & 3.52\textcolor[rgb]{0.733, 0.00, 0.00}{1} & 3.67\textcolor[rgb]{0.00, 0.00, 0.900}{5} & 3.90\textcolor[rgb]{0.00, 0.00, 0.833}{3}
\end{tabular}
\end{ruledtabular}
\end{table}

\subsection{Lattice structure prediction}

Using the optimized Lennard-Jones parameters, we compute lattice energies for all twenty salts in both FCC and BCC lattice arrangements to check whether the experimentally determined crystal structure is correctly favored.
We find that FCC lattice energies are more negative for all twenty salts; this is incorrect for the three BCC salts and correct for the other salts.
However, the gaps between FCC and BCC energies are only 1.6, 1.7, and 2.2~kcal/mol for CsCl, CsBr, and CsI, respectively.
These gaps are narrower than those of all seventeen FCC salts, where the correct lattice was favored by between 2.9 and 22.~kcal/mol, and smaller than the deviations of attained lattice energies from their experimental values.
Lingering deviations from calibration targets for the optimized parameters suggests that a limit on the accuracy of the Lennard-Jones model has been reached; a more realistic model that possibly includes additional parameters is necessary to achieve better agreement with experiments.
This is especially relevant for ions with large electron clouds and high polarizability, such as $\cion{Cs}$ and the larger anions, and may be necessary to correct the lattice structure predictions.

\subsection{Hydration free energies}

To test the transferability of parameters derived using lattice properties, we estimate hydration free energies $\Delta G_\text{hyd}$, an observable that is not used anywhere in their derivation.
Joung and Cheatham~\cite{Joung2008Determination} constructed bicubic surfaces $\Delta G_\text{hyd}^\text{calc}(\sigma,\epsilon)$ that provide single-ion hydration free energies as a function of the ion's $\sigma$ and $\epsilon$ by fitting the results of several hundred thermodynamic integration calculations (note that they used $R_\text{min} = 2^{(1/6)} \sigma$ instead of $\sigma$).
They then combined these surfaces with experimental hydration free energies $\Delta G_\text{hyd}^\text{expt}$ to obtain mappings between $\sigma$ and $\epsilon$ such that $\Delta G_\text{hyd}^\text{calc}(\sigma,\epsilon) = \Delta G_\text{hyd}^\text{expt}$.
However, the bicubic surfaces themselves, which are not influenced by the experimental values, independently constitute a useful and valuable tool because they enable one to estimate the results of hydration free energy calculations for arbitrary Lennard-Jones spheres with monovalent charge to within $\sim0.3$ kcal/mol without performing any simulations.
We employ these surfaces to estimate the calculated hydration free energy for each alkali and halide ion using our Lennard-Jones parameters in TIP3P~\cite{Jorgensen1983Comparison}, TIP4P-Ew~\cite{Horn2004Development}, and SPC/E~\cite{Berendsen1987Missing} water at 298.15~K using a standard state of 1~mol/liter for both gas and solution phases.
As with the original calculations, these estimates incorporate Lorentz-Berthelot mixing rules for ion-water interactions and omit corrections for the liquid-vacuum surface potential~\cite{Asthagiri2003Absolute, Leung2009Ab, Reif2011Computation3}.
Table~\ref{hydrationfreeenergies} compares the estimated hydration free energies to experimental measurements compiled by Schmid et al.~\cite{Schmid2000New}, which also do not account for a liquid-vacuum surface potential.

\begin{table}[htbp]
\caption{\label{hydrationfreeenergies} Negative hydration free energies attained using lattice-derived Lennard-Jones parameters and three water models. Experimental measurements from Schmid et al.~\cite{Schmid2000New} are included for comparison. A temperature of 298.15~K and standard state of 1~mol/liter in both the gas and solution phase applies to all values. Digits colored blue/red indicate where calculated values exceed/fall short of measured values. In each number, the most significant colored digit is shaded to indicate the magnitude of the deviation, with darker shades indicating smaller deviations.}
\begin{ruledtabular}
\begin{tabular}{l d d d d}
 & \multicolumn{3}{c}{Calculated $-\Delta G_\text{hyd}$ (kcal/mol)} & \multicolumn{1}{c}{Measured $-\Delta G_\text{hyd}$} \\
Ion & \multicolumn{1}{c}{TIP3P} & \multicolumn{1}{c}{TIP4P-Ew} & \multicolumn{1}{c}{SPC/E} & \\
\hline
$\cion{Li}$ & 113.\textcolor[rgb]{0.800, 0.00, 0.00}{4} & 10\textcolor[rgb]{0.933, 0.00, 0.00}{6}.\textcolor[rgb]{1.00, 0.00, 0.00}{1} & 11\textcolor[rgb]{0.733, 0.00, 0.00}{2}.\textcolor[rgb]{1.00, 0.00, 0.00}{1} & 113.8 \\
$\cion{Na}$ & 8\textcolor[rgb]{0.700, 0.00, 0.00}{7}.\textcolor[rgb]{1.00, 0.00, 0.00}{6} & 8\textcolor[rgb]{0.867, 0.00, 0.00}{2}.\textcolor[rgb]{1.00, 0.00, 0.00}{5} & 8\textcolor[rgb]{0.767, 0.00, 0.00}{5}.\textcolor[rgb]{1.00, 0.00, 0.00}{6} & 88.7 \\
$\cion{K}$ & 6\textcolor[rgb]{0.700, 0.00, 0.00}{9}.\textcolor[rgb]{1.00, 0.00, 0.00}{7} & 6\textcolor[rgb]{0.833, 0.00, 0.00}{5}.\textcolor[rgb]{1.00, 0.00, 0.00}{8} & 6\textcolor[rgb]{0.800, 0.00, 0.00}{7}.\textcolor[rgb]{1.00, 0.00, 0.00}{5} & 71.2 \\
$\cion{Rb}$ & 65.\textcolor[rgb]{0.867, 0.00, 0.00}{4} & 6\textcolor[rgb]{0.800, 0.00, 0.00}{2}.\textcolor[rgb]{1.00, 0.00, 0.00}{1} & 6\textcolor[rgb]{0.767, 0.00, 0.00}{3}.\textcolor[rgb]{1.00, 0.00, 0.00}{4} & 66.0 \\
$\cion{Cs}$ & 6\textcolor[rgb]{0.00, 0.00, 0.700}{1}.\textcolor[rgb]{0.00, 0.00, 1.00}{5} & 5\textcolor[rgb]{0.733, 0.00, 0.00}{8}.\textcolor[rgb]{1.00, 0.00, 0.00}{9} & 59.\textcolor[rgb]{0.833, 0.00, 0.00}{9} & 60.5 \\
\hline
$\aion{F}$ & 11\textcolor[rgb]{0.700, 0.00, 0.00}{8}.\textcolor[rgb]{1.00, 0.00, 0.00}{7} & 12\textcolor[rgb]{0.00, 0.00, 0.767}{2}.\textcolor[rgb]{0.00, 0.00, 1.00}{4} & 12\textcolor[rgb]{0.00, 0.00, 0.800}{3}.\textcolor[rgb]{0.00, 0.00, 1.00}{8} & 119.7 \\
$\aion{Cl}$ & 8\textcolor[rgb]{0.700, 0.00, 0.00}{7}.\textcolor[rgb]{1.00, 0.00, 0.00}{9} & 90.\textcolor[rgb]{0.00, 0.00, 0.967}{0} & 9\textcolor[rgb]{0.00, 0.00, 0.700}{0}.\textcolor[rgb]{0.00, 0.00, 1.00}{2} & 89.1 \\
$\aion{Br}$ & 8\textcolor[rgb]{0.700, 0.00, 0.00}{1}.\textcolor[rgb]{1.00, 0.00, 0.00}{5} & 83.\textcolor[rgb]{0.00, 0.00, 0.833}{2} & 83.\textcolor[rgb]{0.00, 0.00, 0.833}{2} & 82.7 \\
$\aion{I}$ & 73.\textcolor[rgb]{0.967, 0.00, 0.00}{4} & 74.\textcolor[rgb]{0.00, 0.00, 0.700}{5} & 74.\textcolor[rgb]{0.00, 0.00, 0.700}{4} & 74.3
\end{tabular}
\end{ruledtabular}
\end{table}

The attained RMSDs are 1.0, 4.1, and 2.4~kcal/mol for TIP3P, TIP4P-Ew, and SPC/E, respectively.
Gradients of the bicubic surfaces evaluated at optimized parameter values indicate that hydration free energy is especially sensitive to Lennard-Jones parameters for small ions: across the three water models, the partial derivatives of $-\Delta G_\text{hyd}^\text{calc}$ with respect to $\sigma$ and $\epsilon$ are at least $\left[ 31.9, 77.2 \right]$~kcal/mol for $\cion{Li}$ and $\left[ 38.2, 1513.2 \right]$~kcal/mol for $\aion{F}$.
Therefore, a deviation of the these parameters on the order of $0.1$~\AA{} for $\sigma$ or $0.01$~kcal/mol for $\epsilon$ away from their lattice-calibrated values could significantly degrade the agreement with experimental hydration free energies.
The steepness of these bicubic surfaces is intrinsic to the water models and implies that hydration free energies are a stringent test of the accuracy of Lennard-Jones parameters.
Greater discrepancy between simulation and experiment for TIP4P-Ew compared to the three-site models can be explained by displacement of its negative charge site relative to the oxygen Lennard-Jones center.
This increases the distance from the negative charge of an ion approaching from the oxygen side and accounts for the observed trend of increasing severity of hydration favorability underestimation with decreasing size for the cations~\cite{Grossfield2005Dependence}.
Although caveats regarding assumptions about proton solvation thermodynamics implicit to the experimental values apply~\cite{Grossfield2003Ion}, the general agreement across water models between simulation and experiment on an aqueous system that is completely unlike the crystalline phase used to derive these parameters is evidence of their transferability and validity.

\section{Discussion}

Previous studies have demonstrated the utility of lattice properties as calibration targets.
Peng et al.~\cite{Peng1997Derivation} modified noble gas parameters by adjusting cation sizes and dispersion well depths separately to fit lattice properties and periodic table trends within the framework of a 9-6 van der Waals model with Waldman-Hagler mixing rules.
Gee et al.~\cite{Gee2011KirkwoodBuff} used lattice dimensions to determine dispersion well depths as part of an effort to reproduce experimentally derived Kirkwood-Buff integrals using simulations with SPC/E water.
Joung and Cheatham's thorough study~\cite{Joung2008Determination} fit lattice properties and ion-water binding measurements subject to constraints that maintained accurate hydration free energies, ultimately recommending three completely different sets of parameters for three water models.
They also speculated that it would be possible to use lattice properties as the sole calibration observables in order to achieve water-independent ion parametrization.
The present study confirms this speculation.

The proliferation of alkali and halide ion parameters for nonpolarizable, pairwise additive force fields illustrates the inherent disadvantages of solution properties as calibration targets.
Since solvent model parameters are fixed during these calibrations, any shortcomings and idiosyncrasies of the solvent model become embedded in the resulting parameters.
Therefore, every combination of solvent model and calibration targets engenders its own ion parameters.
In fact, since the calculated solution properties also depend on simulation design choices such as system size, cutoff distance, boundary condition, free energy calculation method, electrostatic approximation method, etc., careful sensitivity analyses must be performed during calibration to prevent the resulting parameters from becoming specific to those choices as well.
For small systems, \emph{ab initio} molecular dynamics and density functional theory~\cite{Varma2011Design, Kelly2007SingleIon} alleviate this problem by explicitly modeling electronic degrees of freedom, providing a transferable way to simulate ions without deriving or choosing a force field.
These calculations may serve as calibration targets for observables that lack definitive experimental measurements.
However, the results are still sensitive to methodological details such as the choice of basis set, pseudopotential, exchange-correlation functional, etc., and simulations of larger systems such as solutions of biopolymers at this level of detail are currently intractable.

The ion parameters presented in this study provide an alternative to the paradigm of solvent-specific customization for use with nonpolarizable, pairwise additive force fields.
Since no choices regarding a solvent model or simulation design were made during their derivation, they are potentially suitable for a broader variety of systems compared to ion parameters built on top of a solvent model.
Lattice-calibrated parameters are immediately applicable to salt crystal and molten salt simulations.
Their independence makes them a reasonable first guess in simulations with solvents besides water that lack customized ion parameters.
As with water, solvent-ion pair potentials determined by the mixing rule may be individually overridden~\cite{Horinek2009Rational, Reif2011Computation4} to attain enhanced accuracy for solvent-specific properties such as solvation structure~\cite{Varma2006Coordination, Whitfield2007Theoretical} and gas phase cluster properties~\cite{Varma2010Multibody}.
Retaining the mixing rule defaults for other pairs preserves the intrinsic character of the ions in their interactions with each other and with other matter.
Encouragingly, the agreement with measured hydration free energies shows that true transferability, where no such tuning is necessary, is not impossible.
In fact, these parameters even make it possible to reverse the direction of dependence by calibrating solvent models while keeping the ion parameters fixed.
In some situations, aqueous simulations may benefit from adopting these parameters despite the abundance of ion parameters tuned for particular water models.
Lattice-calibrated ions are well suited for the ABSINTH implicit solvation model~\cite{Vitalis2009Absinth}, where ion-water geometries are unavailable due to the continuum solvent description and experimental hydration free energies are direct inputs to the model.
They are also justified whenever additional solutes are present and the balance between ion-water and ion-solute interactions is a subject of inquiry; the absence of solvent in their derivation makes them inherently unbiased compared to ion parameters co-derived with a water model.

The general approach of using crystal lattice data to derive ion parameters should remain effective for more demanding problems such as multivalent ions and polarizable force fields.
A wealth of experimental data and techniques from solid state physics are already available, and the combinatorics that allow formation of many different salts from few ions readily lead to overdetermined systems that are desirable in calibrations.
In addition, the periodic nature of crystal lattices makes their calibration observables inherently easier to calculate compared to ones involving dense fluids.
As demonstrated in this study, the use of analytical differentiation and polynomial root solvers make calculation of calibration observables possible with minimal computational effort compared to full equilibrium ensemble simulations, enabling the calculation to be wrapped in a function that is called numerous times by a global minimization routine.
The absence of solvent and symmetry of crystal lattices make this calibration process as simple as possible.

\section{Conclusion}

Crystal lattice properties are sufficient to determine a simultaneous fit of all alkali and halide ion parameters for both the primitive model and Lennard-Jones model with Lorentz-Berthelot mixing rules.
The resulting parameters presented here are transferable, consistent with periodic table trends, and free from the influence of any solvent model.
The success and generality of the method used to derive them suggests that it may be used as a template in future parametrization studies.

\begin{acknowledgments}
This work was supported by National Science Foundation MCB 0718924 and MCB 1121867.
\end{acknowledgments}

\end{document}